\documentclass{article}

\usepackage{epsf}
\usepackage{amssymb}
\usepackage{amsmath}

\begin{document}

\title{On quantum state of entangled photon pairs}
\author{Ruo Peng WANG \\
Department of Physics, Peking University, Beijing 100871, P.R.China \\
Email: rpwang@cis.pku.edu.cn}

\date{\today}
\maketitle

\begin{abstract}
I show that the photon pairs used in experimental tests of quantum non-locality based on Bell's theorem are not in the entangled quantum state. The correct quantum state of the ``entangled'' photon pairs is suggested. Two experiments for testing this quantum state are proposed.


\end{abstract}

\section{introduction}

Quantum non-locality is a controversial topic of quantum theory, and entangled photon pairs played a very important role in experimental tests of quantum non-locality. So far, a number of experimental tests of quantum non-locality based on Bell's theorem \cite{bell} have been carried out \cite{asp,asp1,ou,shih,shih1}. In these experiments entangled photon pairs are produced, and the polarization correlation between entangled photon pairs is measured. Experimental results obtained in these tests are in favor for quantum non-locality. Although it has been pointed out that this polarization correlation is compatible with local realism \cite{sant}, these results are generally considered as direct evidences for the existence of quantum non-locality, and more experiments \cite{str,fon,zhao} showing different kind of quantum non-local correlation are carried out. The entangled photon pairs are used as light source in these experiments, and the interpretation of the experimental results as proofs for quantum non-locality is closely related to the assumption that these photon pairs are in entangled quantum state. 

In this letter, I will show that the same polarization correlation also exists for certain un-entangled photon pairs. And on other hand, based on the consideration of momentum conservation, I will also show that the entangled quantum state is not a correct description for photon pairs used in these experimental tests of quantum non-locality. A correct quantum state for these photon pairs is suggested, and two experiments for testing this quantum state and the quantum non-locality are proposed. 

\section{ polarization correlation between un-entangled photons }
\label{sec:cor}

Let's consider a beam of un-entangled photon pairs with each photon pair containing a left circle polarized photon and a right circle polarized photon. Such a beam of photon pairs can be produced by adjusting polarization state of the photon pairs obtained from the parametric down conversion with a type-II collinear phase matching \cite{kwr}. The quantum state of this un-entangled photon pairs can be expressed as
\begin{equation}
	|\psi \rangle =\frac{1}{2}( b^\dag_v +i b^\dag_h ) 
	( b^\dag_v -i b^\dag_h )|0\rangle = 
	\frac{1}{2}( b^\dag_v b^\dag_v + 
	b^\dag_h b^\dag_h )|0\rangle ,
\end{equation}
where $b^\dag_v$ and $b^\dag_h$ are the creation operator for vertically polarized photon and, respectively, for horizontally polarized photon with the wave vector $\vec k_2$. By using the Coulomb gauge, we may describe the optical field of these photon pairs by the vector potential $\vec A$. The positive frequencies part of the vector potential is related to the photon annihilation operators in the following way  
\begin{equation}\label{inc}
	\vec A^+ = B(b_v \vec e_v + b_h \vec e_h)e^{i\vec k_2 \cdot \vec r}
\end{equation}
where the constant $B$ is given by \cite{mandl}
\begin{equation}
	B= \sqrt{\frac{\mu_0 \hbar c^2}{2 V \omega}}
\end{equation}
with $\mu_0$ the magnetic permeability of the vacuum, $\omega$ the angular frequency of the photons.

\begin{figure}
\begin{center}
\epsfbox{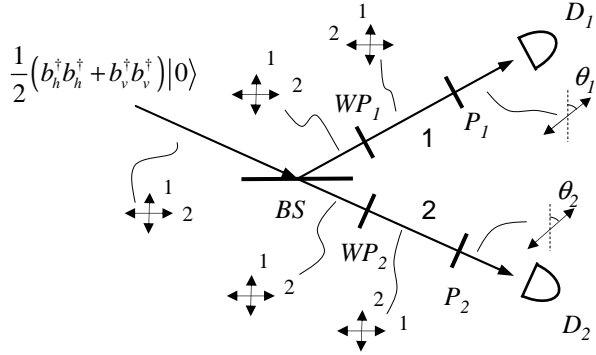} 
\end{center}
\caption{\label{fig1} A schematic illustration of the experimental setup for testing the polarization correlation between un-entangled photon pairs. The polarization states of optical beams are indicated by arrows.}
\end{figure}

We divide the beam of un-entangled photon pairs into two channels by using a half reflecting mirror ($BS$). The wave vectors of photons in these two channels are $\vec k_1$ and $\vec k_2$. A half wave plate ($WP_1$) is placed into the channel 1 to introduce a phase difference of $\pi$ between the vertical and horizontal components of the optical field, and another half wave plate ($WP_2$) is placed into the channel 2 to swap the vertical and horizontal components of the optical field, as shown in FIG. \ref{fig1}. As no incident photons with the wave vector $\vec k_1$ being present, after passing the beam splitter and the wave plates, the optical field becomes \cite{wolf}
\begin{equation}\label{ref_tr}
	\vec A^+ =  \frac{B}{\sqrt{2}}(b_v \vec e_v - b_h \vec e_h)
	e^{i\vec k_1 \cdot \vec r}
	+ \frac{B}{\sqrt{2}}(b_h \vec e_v + b_v \vec e_h)
	e^{i\vec k_2 \cdot \vec r} .
\end{equation}
The photons in the channel 1 and 2 are polarized by Glan-Thompsom linear polarization analyzers $P_1$ and $P_2$, respectively, before being detected by single photon detectors $D_1$ and $D_2$.

To analyze the polarization correlation between photons detected by detectors $D_1$ and $D_2$, we need expressions of optical field at these detectors. Up to a phase factor, we have 
\begin{equation}\label{a1}
	\vec A^+_{\theta_1} =  \frac{B}{\sqrt{2}}
	(b_v \cos \theta_1 - b_h \sin \theta_1)
	\vec e_{\theta_1} 
\end{equation}
and
\begin{equation}\label{a2}
	\vec A^+_{\theta_2} =  \frac{B}{\sqrt{2}}
	(b_h \cos \theta_2 + b_v \sin \theta_2)
	\vec e_{\theta_2}.
\end{equation}
In the above expressions, $\vec e_{\theta_1}$ and $\vec e_{\theta_2}$ are vectors of unit of the transmission axes of the analyzer $P_1$, and respectively, $P_2$. The relations (\ref{a1}) and (\ref{a2}) can be rewritten in the following form 
\begin{equation}
	\vec A^+_{\theta_1} =  \frac{B}{\sqrt{2}}	b_{\theta_1}\vec e_{\theta_1},\;
	\vec A^+_{\theta_2} =  \frac{B}{\sqrt{2}}
	b_{\theta_2}\vec e_{\theta_2} 
\end{equation}
where 
\begin{equation}
	b_{\theta_1} = b_v \cos \theta_1 - b_h \sin \theta_1,\;
	b_{\theta_2} = b_h \cos \theta_2 + b_v \sin \theta_2
\end{equation}
are the annihilation operators for the photon polarized in the direction $\vec e_{\theta_1}$ with the wave vector $\vec k_1$ and, respectively, for the photon polarized in the direction $\vec e_{\theta_2}$ with the wave vector $\vec k_2$.

The coincidence counting rate $C(\theta_1,\theta_2)$ is proportional to the probability of annihilating simultaneously one photon polarized in the direction $\vec e_{\theta_1}$ at the detector $D_1$ and one photon polarized in the direction $\vec e_{\theta_2}$ at the detector $D_2$. This probability is, in its turn, proportional the matrix element
\begin{equation}
\begin{split}
	\langle \psi | b^\dag_{\theta_2}b^\dag_{\theta_1}
	b_{\theta 1}b_{\theta 2} |\psi \rangle & =
	\langle \psi | ( b^\dag_v \cos \theta_1 - b^\dag_h \sin \theta_1)
	(b^\dag_h \cos \theta_2 + b^\dag_v \sin \theta_2) \\
	&\quad \times( b_v \cos \theta_1 - b_h \sin \theta_1)
	(b_h \cos \theta_2 + b_v \sin \theta_2)
	|\psi \rangle \\
	& = \frac{1}{4}\sin^2 (\theta_1 - \theta_2).
\end{split}
\end{equation}
Therefore we have
\begin{equation}\label{cor}
	C(\theta_1,\theta_2)=C_M \sin^2 (\theta_1 - \theta_2),
\end{equation}
where $C_M$ this the maxim counting rate that occurs at $\theta_1-\theta_2 = \pm \pi/2$. This correlation is just the same as in the case of entangled photon pairs \cite{shih1,kwr}, and the Bell's inequality is apparently violated also in the above discussed experiment. The violation of the Bell's inequality in this case is only apparent, because a condition for obtaining the Bell's inequality is not fulfilled here. Namely, in the experiment described in this paper, both photons of a photon pair could be detected in one single channel, with the wave vector $\vec k_1$ or $\vec k_2$. Thus, even though the same polarization correlation as in the case of entangled photon pairs is realized in this experiment, no quantum non-locality is involved here. Indeed, after annihilating of a linear polarized photon at the detector $D_1$, the quantum state becomes
\begin{equation}
	|\psi^\prime \rangle = 
       b_{\theta_1} |\psi \rangle 
       =\frac{1}{\sqrt{2}}( \cos\theta_1 b^\dag_v - 
	\sin\theta_1 b^\dag_h )|0\rangle .
\end{equation}
The photon in the quantum state $|\psi^\prime \rangle $ is not spatially separated from the detector $D_1$ where another photon was annihilated. The probability as the photon in the quantum state $|\psi^\prime \rangle $ being detected at $D_1$ is
\begin{equation}
	\langle\psi^\prime | b^\dag_{\theta_1} b_{\theta_1} 
       |\psi^\prime \rangle = \frac{1}{2}.
\end{equation}

\section{quantum state of entangled photon pairs}
\label{sec:state}

Now we arrive at a very important point: the observation of a photon polarization correlation as expressed in the relation (\ref{cor}) does not form a sufficient evidence for quantum non-locality. Besides this correlation, one must exclude the possibility of detecting both of the photons in one single channel to ensure that the phenomena of quantum non-locality are really observed.

An apparently non-local correlation as expressed in the relation (\ref{cor}) are observed experimentally for the entangled photon pairs generated in the parametric down conversion nonlinear optical processes with a type-II phase matching. We have to see if the possibility of detecting both of the photons in one single channel can be ruled out in the experiment so that one can really interpret the experimental results as an experimental evidence for quantum non-locality. So far the following quantum state is used to describe this photon pair\cite{kwr}:
\begin{equation}\label{ent}
	|\psi_e \rangle = \frac{1}{\sqrt{2}}( b^\dag_{v1} b^\dag_{h2}
	- b^\dag_{v2} b^\dag_{h1})|0 \rangle,
\end{equation}
where $ b^\dag_{h1},b^\dag_{v1}, b^\dag_{h2}$ and $b^\dag_{v2}$ are creation operators for horizontally polarized photons with the wave vector $\vec k_1$,  
for vertically polarized photons with the wave vector $\vec k_1$, for horizontally polarized photons with the wave vector $\vec k_2$, and for vertically polarized photons with the wave vector $\vec k_2$, respectively. By using the corresponding photon annihilation operators, we may write the optical field as
\begin{equation}\label{a_e}
	\vec A^+_e =  B(b_{v1} \vec e_v + b_{h1} \vec e_h)
	e^{i\vec k_1 \cdot \vec r}
	+ B(b_{v2} \vec e_v + b_{h2} \vec e_h)
	e^{i\vec k_2 \cdot \vec r} .
\end{equation}
It is easy to verify that there exists a correlation as described by the relation (\ref{cor}) between the polarization of the photon pairs in the quantum state $|\psi_e \rangle $.

The expression (\ref{ent}) for the quantum state of the entangled photon pairs was derived by taking into consideration of the energy and the momentum conservation in the parametric down conversion process. The energy and the momentum of a photon pair in the state $|\psi_e \rangle $ is equal to the energy and the momentum of the incident photon from which the photon pair is generated. But there is a problem in this argument, namely the parametric down conversion process takes place in nonlinear optical crystals, so one must take the photons and the crystal as one single system when the condition of the energy and the momentum conservation is applied. The macroscopic properties of nonlinear optical crystals do not change in the parametric down conversion process. Thus the expectation values of the energy and the momentum of crystals are the same before and after the photon pairs' generation. Therefore the expectation values of energy and momentum of a photon pair must equal to the expectation values of energy and momentum of incident photon. This condition is satisfied by the quantum state $|\psi_e \rangle $. On other hand, photons are coupled with optical phonons with same momentum when passing through crystals, and there exist exchanges of energy and momentum between photons and crystals through the photon-phonon interaction during photons' propagation in crystals. One may neglect the energy exchange between photons and crystals because the energies of phonons are much smaller than that of photons with same momentum, but the momentum exchange between photons and crystals must be taken into consideration. This exchange of momentum causes fluctuations in momenta of crystals and generated photon pairs. But the state $|\psi_e \rangle $ is an eigen-state of the photon pairs' momentum, that means there is not momentum fluctuations for photon pairs in the state $|\psi_e \rangle $. Therefore the generated photon pair can not be in this quantum state.

According to the calculation made before, the correlation between the polarization of the photon pairs as described by the relation (\ref{cor}) does exist also for photon pairs in the following quantum state
\begin{equation}\label{f_u}
	|\psi_u \rangle =\frac{1}{2}( b^\dag_{1} +i b^\dag_{2}) 
	( b^\dag_{1} -i b^\dag_{2})
	|0\rangle =  
	\frac{1}{2}( b^\dag_{1} b^\dag_{1} + 
	b^\dag_{2} b^\dag_{2})|0\rangle, 
\end{equation}	
with the optical field given by
\begin{equation}\label{a_u}
	\vec A^+_u = \frac{B}{\sqrt{2}}(b_1 \vec e_v - b_2 \vec e_h)
	e^{i\vec k_1 \cdot \vec r}
	+ \frac{B}{\sqrt{2}}(b_2 \vec e_v + b_1 \vec e_h)
	e^{i\vec k_2 \cdot \vec r} .
\end{equation}
The quantum state $|\psi_u \rangle $ is an un-entangled state. In fact, we can write $|\psi_u \rangle $ as
\begin{equation}
	|\psi_u \rangle = |\psi_a\rangle \otimes |\psi_b\rangle, 
\end{equation}	
with
\begin{equation}
	|\psi_a\rangle = b^\dag_a |0\rangle,\;
	|\psi_b\rangle = b^\dag_b |0\rangle, 
\end{equation}
where
\begin{equation}
	b^\dag_a =\frac{1}{\sqrt{2}}( b^\dag_{1} +i b^\dag_{2}) ,\;
	b^\dag_b =\frac{1}{\sqrt{2}}( b^\dag_{1} -i b^\dag_{2}),\;
	[ b^\dag_a,\, b_b]=0. 
\end{equation}

The quantum state $|\psi_u \rangle $ is not an eigen-state of momentum, but photon pairs in the quantum states $|\psi_e \rangle $ and $|\psi_u \rangle $ have the same expectation values of the energy and the momentum. So the quantum state $|\psi_u \rangle $ satisfies the energy and momentum conservation requirement. In conclusion, the quantum state $|\psi_u \rangle $ could be a correct description for photon pairs generated in the parametric down conversion nonlinear optical processes with a type-II phase matching. One may observe that for photon pairs in the quantum state $|\psi_u \rangle $, both photons could be detected in one single channel. Therefore, by applying the conditions of the energy and momentum conservation, we can not rule out the possibility of detecting both photons in one single channel. But, instead, we find that the photon pairs can not be in the entangled state. The apparently non-local correlation between the polarization of photon pairs produced in the parametric down conversion nonlinear optical processes, that were believed being in entangled states, is not a proof for the existence of the quantum non-locality, but just a necessary evidence for the fact that photon pairs are in the quantum state $|\psi_u \rangle $. 

The same conclusion holds also for photon pairs emitted in a radiative atomic cascade of calcium \cite{asp,asp1}. In that process, electrons which emit two photons in a radiative cascade are well confined within the ions of calcium. The uncertainty in the momentum of electrons implies that the photon pairs can not be in the entangled state which has a well defined momentum. The quantum state for photon pairs generated in this process could be expressed as
\begin{equation}\label{f_up}
	|\psi^\prime _u \rangle = \frac{1}{\sqrt{2}}
	(b^\dag_{\omega_1h} b^\dag_{\omega_2h}
	+b^\dag_{\omega_1v} b^\dag_{\omega_2v}) |0\rangle , 
\end{equation}	
where $b^\dag_{\omega_1h},b^\dag_{\omega_1v}$ are creation operators for horizontally, and respectively, vertically polarized photons of circle frequency $\omega_1$, and $b^\dag_{\omega_2h},b^\dag_{\omega_2v}$ are the same operators for photons of circle frequency $\omega_2$.

The optical field in the channel 1 is given by
\begin{equation}\label{a_1}
	\vec A^+_1 = g_{11}(b_{\omega_1v} \vec e_v + b_{\omega_1h}
	\vec e_h) e^{i c \omega_1 \vec n_1 \cdot \vec r}
	+ g_{12}(b_{\omega_2v} \vec e_v + b_{\omega_2h}
	\vec e_h) e^{i c \omega_2 \vec n_1 \cdot \vec r}
\end{equation}
and
\begin{equation}\label{a_2}
	\vec A^+_2 = g_{21}(b_{\omega_1v} \vec e_v + b_{\omega_1h}
	\vec e_h) e^{i c \omega_1 \vec n_1 \cdot \vec r}
	+ g_{22}(b_{\omega_2v} \vec e_v + b_{\omega_2h}
	\vec e_h) e^{i c \omega_2 \vec n_1 \cdot \vec r}
\end{equation}
in the channel 2, where $\vec n_1, \vec n_2$ are vectors of unity that indicate the propagation directions of the channels 1 and 2, and $g_{11}, g_{12}, g_{21}, g_{22}$ are coefficients that depend on the geometry of the experiment setup. We can now calculate the probability of annihilating simultaneously a photon of circle frequency $\omega_1$ polarized in the direction $\vec e_{\theta_1}$ in the channel 1 and a photon of circle frequency $\omega_2$ polarized in the direction $\vec e_{\theta_2}$ in the channel 2. By using the expressions (\ref{f_up}), (\ref{a_1}) and (\ref{a_2}), we find
\begin{equation}\label{cor_p}
	C_{12}(\theta_1,\theta_2)=C_M^\prime \cos^2 (\theta_1 - \theta_2).
\end{equation}
This is exactly the correlation observed in experimental tests carried out by Aspect {\it et al}. \cite{asp,asp1}. Again, no quantum non-locality is evolved,
and both of the photons can be detected in one single channel also in these tests.

\section{proposal for experimental tests}
\label{sec:test}

\begin{figure}
\begin{center}
\epsfbox{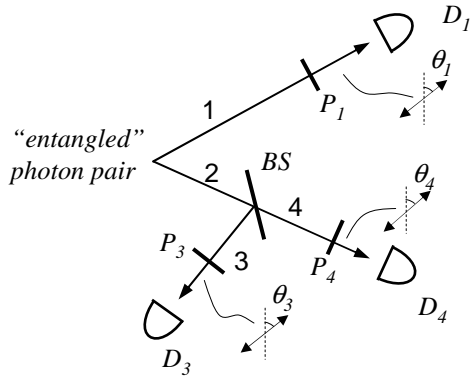} 
\end{center}
\caption{\label{fig2} A schematic illustration of the experimental setup for testing the possibility of detecting both photons from one ``entangled'' photon pair in one single channel. The polarization states of optical beams are indicated by arrows. }
\end{figure}

Direct experimental tests on the possibility of detecting both photons in one single channel and on the coherence of photons in different channels could make our conclusions on the quantum state of ``entangled'' photons pairs and quantum non-locality more convincing. An experimental test on the possibility of detecting both of the photons in one single channel can be done by using an experimental setup shown in FIG. {\ref{fig2}}. This setup is quite similar to that used for tests of quantum non-locality based on Bell's theorem. But a half reflecting mirror is inserted now into the channel 2 to split it into the channels 3 and 4, and the coincidence counting rate between the channels 3 and 4 is measured. This coincidence counting rate is proportional to the polarization correlation between photons in the channel 3 and in the channel 4. An anti-coincidence condition with the signal from the channel 1 can also be applied to ensure that this coincidence counting rate is not from other sources. By using the expressions (\ref{f_u}),  (\ref{a_u}), (\ref{ent}) and (\ref{a_e}), one can easily verify that the following relation holds for this coincidence counting rate 
\begin{equation}\label{cor23}
	C(\theta_{3},\theta_4)=C_M^{\prime\prime} \cos^2 (\theta_{3} - \theta_4),
\end{equation}
if the photon pairs are in the quantum state $|\psi_u \rangle $, and
\begin{equation}\label{cor23e}
	C(\theta_{3},\theta_4) \equiv 0,
\end{equation}
if the photon pairs are in the quantum state $|\psi_e \rangle $.

\begin{figure}
\begin{center}
\epsfbox{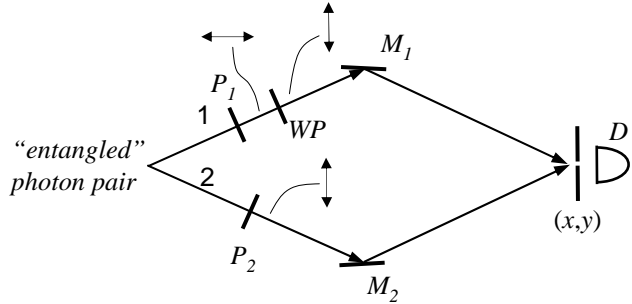} 
\end{center}
\caption{\label{fig3} A schematic illustration of the experimental setup for testing the coherence between photons from one ``entangled'' photon pair in different channels. The polarization states of optical beams are indicated by arrows. }
\end{figure}

The coherence of photons in different channels can also be used for testing the quantum state of ``entangled'' photon pairs. An experimental setup for such a test is schematically illustrated in FIG. \ref{fig3}. The linear polarizers $P_1$ and $P_2$ are inserted into the beams of ``entangled'' photon pairs,
generated in the parametric down conversion nonlinear optical processes with a type-II phase matching, in such a way, so that the photon in the channel 1 becomes horizontally polarized, while the photon in the channel 2 is polarized vertically. The polarization of the beam 1 is changed to vertical later by the half wave plate $WP$. Both beams are reflected by the mirrors $M_1$ and $M_2$ to overlap each other. The single photon counting rate $I(x,y)$ in the $(x,y)$ plan as a function of the coordinates $(x,y)$ is measured by using the single photon detector $D$. Let $f_1(x,y)\vec e_v$ be the distribution of the vector potential of the beam 1 in the plan $(x,y)$, and $f_2(x,y)\vec e_v$ the distribution of the vector potential of the beam 2 . If the photon pair were in the quantum state $|\psi_e \rangle $, we have \begin{equation}\label{coh1}
	\vec A^+(x,y) = b_{h1} f_1(x,y) \vec e_v +  b_{v2} f_2(x,y) \vec e_v .
\end{equation}

The single photon counting rate $I(x,y)$ is proportional to the matrix element
\begin{equation} 
	I(x,y) \propto \langle \psi_e | \vec A^{+\dag}(x,y) \cdot
       \vec A^+(x,y)  |\psi_e \rangle .
\end{equation}
By using the relation (\ref{coh1}) and the expression (\ref{ent}) for $|\psi_e \rangle $, we find, for photon pairs in the quantum state $|\psi_e \rangle $,
\begin{equation} 
       	I(x,y) \propto |f_1(x,y)|^2 + |f_2(x,y)|^2,
\end{equation}
so no interference occurs. But if the photon pairs were in the quantum state $|\psi_u \rangle $, then according to Eq.(\ref{a_u}), we have
\begin{equation}\label{coh2}
	\vec A^+(x,y) = b_{2} f_1(x,y) \vec e_v +  b_{2} f_2(x,y) \vec e_v .
\end{equation}
And in this case the counting rate becomes
\begin{equation} 
       	I(x,y) \propto |f_1(x,y) + f_2(x,y)|^2,
\end{equation}
that means the beam 1  and beam 2 prepared in the way described above are coherent.

\section{discussion}
\label{sec:con}

I have shown that the correlations of photons' polarization observed in ``entangled'' photon pairs generated in the parametric down conversion 
nonlinear optical processes with a type-II phase matching and in a radiative atomic cascade of calcium are not proofs for quantum non-locality.
Instead, they are necessary evidences for the fact that ``entangled'' photon pairs are in the un-entangled states $|\psi_u \rangle $ or $|\psi_u^\prime \rangle $. 
According to the expression (\ref{a_u}) for the vector potential, 
in the case of ``entangled'' photon pairs generated in the parametric down 
conversion nonlinear optical processes with a type-II phase matching,
we may express the operators $b^\dag_{1}$ and $b^\dag_{2}$ in terms of operators $b^\dag_{v1}$, $b^\dag_{v2}$, $b^\dag_{h1}$ and $b^\dag_{h2}$:
\begin{equation} 
	b^\dag_{1}= \frac{1}{\sqrt{2}}(b^\dag_{v1} + b^\dag_{h2}),\;
	b^\dag_{2}= \frac{1}{\sqrt{2}}(b^\dag_{v2} - b^\dag_{h1}) .
\end{equation}
We have then
\begin{equation}\label{f_u2}
	|\psi_u \rangle =\frac{1}{\sqrt{2}} |\psi_e \rangle 
	+ \frac{1}{4}( b^\dag_{v1} b^\dag_{v1} + 
	b^\dag_{v2} b^\dag_{v2}+ b^\dag_{h1} b^\dag_{h1} + 
	b^\dag_{h2} b^\dag_{h2})|0\rangle.
\end{equation}	
As the component $( b^\dag_{v1} b^\dag_{v1} + b^\dag_{v2} b^\dag_{v2}+ b^\dag_{h1} b^\dag_{h1} + b^\dag_{h2} b^\dag_{h2})|0\rangle $ in $|\psi_u \rangle $ has no contribution to the coincidence counting rate between the signals from the channel 1 and the channel 2, all apparently non-local correlations that were believed as specificity of photon pairs in the entangled state $|\psi_e \rangle $, occur also in the case of un-entangled photon pairs in the state $|\psi_u \rangle $. The same conclusion holds also for 
``entangled'' photon pairs generated in a radiative atomic cascade of calcium
Therefore no physical phenomena that necessitate introducing quantum non-locality for their explanation are really observed.

Einstein, Podolsky, Rosen (EPR) and Bohm had put the completeness of quantum mechanics in contradiction to the relativistic causality by supposing the existence of particle pairs in entangled quantum states \cite{epr,bohm}. 
But till now, such a contradiction did not occur, because no particle pairs in entangled quantum states had been produced. Can particle pairs in entangled quantum states be generated ever? It is most likely not. Due to the interaction with the source of particle pairs, it should be impossible for the produced 
particle pair with different momenta to be in a quantum state with well defined momentum, such as the entangled quantum state. This observation is consistent with Santos's suggestion \cite{sant} that only quantum states which do not contradict with locality requirement are physical states. From this point of view, the EPR paradox is just a spectacular illustration of the restriction on quantum states imposed by locality requirement.

\end{document}